\documentclass[a4paper]{jpconf}

\def\be{\begin{eqnarray}}
\def\ee{\end{eqnarray}}
\def\lsim{\;\raise0.3ex\hbox{$<$\kern-0.75em\raise-1.1ex\hbox{$\sim$}}\;}
\def\gsim{\;\raise0.3ex\hbox{$>$\kern-0.75em\raise-1.1ex\hbox{$\sim$}}\;}

\newcommand{\beq}{\begin{eqnarray}}
\newcommand{\eeq}{\end{eqnarray}}
\def\lsim{\;\raise0.3ex\hbox{$<$\kern-0.75em\raise-1.1ex\hbox{$\sim$}}\;}
\def\gsim{\;\raise0.3ex\hbox{$>$\kern-0.75em\raise-1.1ex\hbox{$\sim$}}\;}

\def \hcm {\hbox {\ifmmode $ atom cm$^{-2}\else atom cm$^{-2}$\fi}}

\def \AA {\AA}

\def\approxgt{\mathrel{\hbox{\rlap{\lower.55ex \hbox {$\sim$}}
        \kern-.3em \raise.4ex \hbox{$>$}}}}
\def\approxlt{\mathrel{\hbox{\rlap{\lower.55ex \hbox {$\sim$}}
        \kern-.3em \raise.4ex \hbox{$<$}}}}

\usepackage{graphicx}
\usepackage{cite}
\begin{document}
\title{Simulation of collisionless shocks in plasmas with high metallicity}
\author{J A Kropotina$^1$, K P Levenfish$^1$ and A M Bykov$^1$}
\address{$^1$ The Ioffe Institute}
\ead{juliett.k@gmail.com}
\begin{abstract}
The results of hybrid simulation  of low-beta supercritical quasi-parallel shocks in high metallicity plasma are presented. The structure of upstream and downstream turbulence is addressed and velocities of the corresponding scattering centers are derived. It is shown that independently of their chemical composition the shocks experience self-reformation process. However, the period of self-reformation as well as the wave spectrum is greatly affected by the presence of substantial admixture of weakly charged heavy ions. Also the downstream magnetic field amplification is stronger for the high metallicity case.
\end{abstract}
\section{Introduction}
\label{s_intro} Collisionless shocks, appearing on scales much less than the Coulomb mean free path, are ubiquitous in astrophysical plasma. They are governed by self-consistent wave-particle interactions and act as effective particles accelerators \cite{Krymskii77} as well as generators of amplified magnetic field via various plasma instabilities (see, e.g., \cite{treumann09,Marcowith_2016}). The shock transition is produced 
either by electron resistivity for subcritical shocks (with Mach number lower than a few units) or by the ions instabilities caused by the shock-reflected ions for the supercritical ones. (\cite{Sagdeev66,tk71,treumann09, Kennel_ea85, Lembege_ea04,Burgess_ea05}). 
The latter are governed primordially by ions and therefore can be successfully simulated by means of hybrid codes, which treat ions kinetically while assuming electrons to be a collisionless neutralizing fluid. 

One of the prominent supercritical shocks features is the appearance of \textit{self-reformation} above the so-called second critical Mach number \cite{Krasnoselskikh2002}. This process involves quasi-periodical steepening and broadening of the shock ramp, caused by reflection of a minor ions population during the steepened shock phase with its subsequent accumulation at close upstream. The relative streaming of the incoming and reflected ions leads to nonlinear wave generation and formation of a new ramp ahead of the previous. As a result, in the downstream reference frame the shock front stays at nearly constant position during most of the reformation cycle, and then rapidly jumps ahead, in contrast to the uniform front movement predicted by the simple hydrodynamics. The self-reformation appears for both quasi-parallel (the magnetic field inclination angle $\theta < 45^o$) and quasi-perpendicular ($\theta > 45^o$) shocks. 

Due to the self-consistent interplay between shock steepening and ion reflection the shock reformation is closely related to the ions acceleration process. One of the most efficient mechanisms for charged particles acceleration to relativistic energies is diffusive shock acceleration (DSA). It consists of the ions repeated reflections upstream and downstream the shock, which leads to energy gain due to the difference of scattering centers velocities. The key point is that the particles need to be substantially pre-accelerated (\textit{injected}) to overcome the cross-shock potential and start scattering at both sides of the shock front. According to recent numerical and observational works \cite{ Sundberg16,Caprioli2015},
the particles which subsequently succeed to gain supra-thermal energy and enter the DSA are extracted from the initially reflected population. This suggests that the ions injection efficiency varies during the reformation cycle together with the reflection percentage.

The self-reformation process for the pure hydrogen shocks have been thoroughly investigated by means of both numerical simulations
\cite{Lembege1992,Thomas1990JGR, Winske1990JGR, Caprioli2015} and in-situ solar wind observations \cite{Lobzin2008}, the reformation period being of order a few upstream proton gyrofrequencies, depending on shock parameters. At the same time in supernovae reverse shocks the plasma composition differs dramatically. The ejecta of supernova remnants is highly enriched by heavy elements, so that the shock dynamics changes, as well as characteristic temporal and spatial scales of the related processes. 

In this paper we investigate how the substantial admixture of weakly charged heavy ions affects the shock reformation process. We consider a quasi-parallel shock in cold magnetized plasma, consisting of two sorts ions with equal densities and substantially different gyrofrequencies. It is found that in this case shock reformation process is governed by both sorts of ions, and hence two characteristic reformation periods appear, differing in $A/Z$ times (where $A$ and $Z$ are the ions mass and charge number respectively). At the same time the power in long-wave fluctuations, produced by heavy ions, is several times higher.
Accordingly, a stronger downstream magnetic field grows in high metallicity shock, indicating the possibility of effective magnetic field amplification in multi-component plasma. 

The other goal is to check from the first principles the analytical predictions about  velocity of the downstream and upstream scattering centers (i.e. wave fronts). This can help to build semi-analytical and Monte-Carlo models of the DSA process.

The paper is organized as follows: in section \ref{sim} the simulation technique and setup are briefly described, while the results are given in section \ref{res}.
  
\section{Simulations}
\label{sim}
The shocks are simulated by means of the 3d second-order accurate divergence-conserving hybrid code \textit{Maximus} (see \cite{Kropotina15, Kropotina16, Kropotina19} for the thorough code description). Hybrid code is a common type of kinetic particle-in-cell codes, which treats electrons as collisionless neutralizing fluid, meanwhile solving the full collisionless Vlasov-Maxwell equations system for ions (see, e.g., \cite{Matthews94, WinskeO96, Lipatov}). The equations are solved on a cartesian grid with cells of constant fields and currents. The characteristics (the so-called \textit{macroparticles}, each of them representing a co-moving ensemble of real ones) are introduced for the Vlasov equation. The quadratic current interpolation ensures second order accuracy, while the total variation diminishing Godunov-based algorithm is implemented for the divergence-free calculation of electromagnetic fields. The usage of hybrid technique allows to make simulations on ionic scales, which far exceed the electronic ones. Therefore greater simulations boxes and times are available with the given computer resources.

The shock is initiated via the 'rigid piston` method, when the superalfvenic flow reflects from a conducting wall and forms a shock due to the ion beam instability. Therefore the simulation takes place in the downstream reference frame, and the shock front moves against the incoming flow. A non-relativistic shock properties are fully determined by the three upstream dimensionless parameters: the Alfv\'en Mach number $M_a$, the ratio of thermal to magnetic pressure $\beta$ and the inclination angle $\theta$. The latter is expected to vary along the shock front due to long-scale turbulence, so we choose the arbitrary value $\theta = 10^o$. The additional simulations with $\theta = 0^o$ and $\theta = 30^o$ showed qualitatively similar results. We focus on the quasi-parallel configuration, which is preferential for ions acceleration via the 1 order Fermi mechanism (diffusive shock acceleration) \cite{Caprioli14a}. At the same time the former two parameters
depend on the magnetic field in the ejecta, which is actually unknown. Here we substitute the rather high value of $B_0 \sim 1mG$, found from the X-ray observational data for the Cas A supernova remnant in \cite{Sato18}.  This leads to relatively low $M_a \sim 10$ and $\beta \sim 10^{-3}$. It should be noted that high Mach numbers require small timesteps, therefore being computationally expensive. On the other hand, extremely low betae can lead to numerical challenges due to the Alfv'en decay instability (see, e.g., \cite{Amano14}).
So, partly for numerical reasons, the $M_a=10$ and $\beta = 0.002$ have been chosen. In all setups the incoming flow propagates along $-x$ direction, the magnetic field lies in $x-z$ plane, and the reflecting wall is at $x=0$. The simulation box spannes $8000\times1\times600$ cells of size one cubic ion inertial length $l_i = c/\omega_{pi}$, where $\omega_{pi}$ is proton plasma frequency. Hereafter all lengths are normalized to $l_i$, times to the proton gyrofrequency $\Omega = eB_0/mc$, and magnetic fields to the far upstream value $B_0$.

To investigate the effect of the heavy ions admixture two shocks with similar parameters but different chemical composition are considered. The first one is a well investigated pure hydrogen shock, while in the second one the heavy ions are fiducially represented by the 50\% (by mass) admixture of O(+1). Therefore the shock is governed by the two ions, which initially carry the same energy and which gyrofrequencies and gyroradii  differ in 16 times. The resulting picture is described below.

\section{Results}
\label{res}
\begin{figure}[h]
\includegraphics[width=0.9\linewidth]{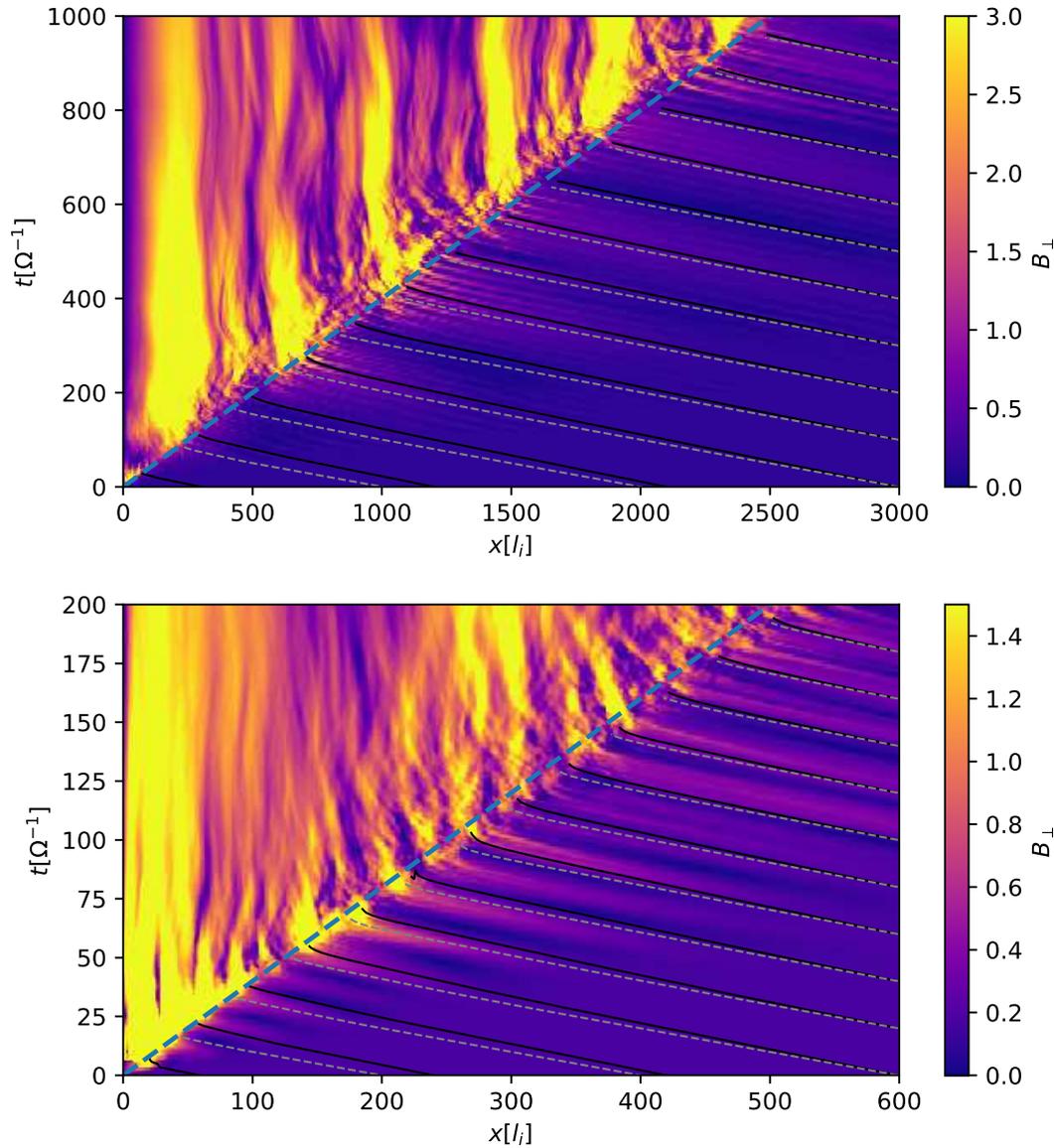}
\caption{Temporal evolution of perpendicular magnetic field for the shock with 50\% O(+1) admixture (top panel) compared to the same for the pure hydrogen shock (bottom panel). The analytically found characteristics for upstream resonant streaming instability waves are shown by solid black lines, while the characteristics for perturbations moving with the upstream plasma are given by dashed gray lines for comparison. Front position is indicated by the blue dashed line.}
\label{time_evolution}
\end{figure}

The evolution of magnetic field component, perpendicular to the shock normal, is colormapped in fig. \ref{time_evolution}. The bottom panel corresponds to the pure hydrogen shock, while the top one to the shock with 50\% O(+1) admixture. It can be seen that in both cases the shock propagates with approximately the same averaged velocity $V_{sh} \approx 2.5 V_a$ (see the blue dashed lines indicating the shock position). Also the averaged density compression $r \equiv \rho_{down}/\rho_{up}$ (not shown) is very close for various compositions. So it can be concluded that the macroscopic shock parameters are not greatly affected by the substantial admixture of weakly charged heavy ions. It should be noted that the shock compression ratio is $r \approx 5$, which is greater than the standart Rankine-Hugoniot prediction for a strong shock ($r = 4$). This appears due to the shock modification by substantial pressure of accelerated particles (see, e.g., \cite{McKenzieVolk82}). Accordingly, the shock front velocity in the downstream frame is less than the hydrodynamical value $V_{sh,h} \approx 3.3 V_a$ for the upstream $M_a = 10$.

The waves in the precursor are in both cases produced by the resonant streaming instability, as can be seen from their properties and spectra. In good agreement with analytical predictions \cite{Bell78a, McKenzieVolk82}, the generated waves are of the Alfv\'en type and propagate with positive velocity $V = V_a$ in the upstream reference frame (see the overplotted analytical characteristics). The characteristics for the hypothetical perturbations co-moving with the upstream plasma are plotted by dashed gray lines for comparison. It is obvious that they do not agree with the simulation results. However, it should be noted that kinetic codes due to their high resource-intensity are not able to trace particles acceleration and magnetic field amplification up to scales comparable with the Supernovae remnants lifetimes and sizes. Meanwhile the development of the cosmic rays-induced large-amplitude turbulence is likely to break the quasi-linear approximation, so that the scattering centers velocity may differ from $V_a$. The analytical \cite{Caprioli2012} and Monte-Carlo \cite{Bykov14} models have been developed for this case.

At the same time the downstream fields are much less ordered. However, it can be seen that the downstream field perturbations are on average steady in plasma frame and damp with time. So it can be concluded, that independently of  plasma composition the upstream scattering centers move with the Alfv\'en velocity against the incoming flow, while the downstream ones are approximately at rest in the plasma frame.

Meanwhile there persist dramatic differences in the magnetic fields for various plasma compositions. The first obvious fact is the presence of approximately twice as high magnetic fields downstream the oxygen-enriched shock as those for the pure hydrogen one (note different color scales in fig. \ref{time_evolution}). This indicates the possibility of more effective magnetic field amplification in plasma with high metallicity.

The microphysics of shock reformation seems to be the same for both cases, but it can be easier captured for the oxygen-enriched shock (upper panel of fig. \ref{time_evolution}) due to larger periods and amplitudes. It can be seen that  the higher values of $B_z$ at front (i.e. the phase of steepened shock ramp) appear immediately after the arrival of the upstream wave, produced by the reflected ions in the shock precursor and convected back to the shock front. The shock steepening in turn leads to the accumulation of the reflected ions and growing of a magnetic field bump in the close upstream. Simultaneously the previous steepened ramp damps while the bump grows up and becomes a new ramp. This is the broadened shock phase, which appears to be several times longer than the steepened one (see also \cite{Caprioli2015}, where the pure hydrogen shock is considered to spend 25\% of cycle in the steepened state). 

From the bottom panel of figure \ref{time_evolution} the reformation period for pure protons can be estimated as roughly $10 \Omega^{-1}$. The same ``short'' reformation periods persist in the case of oxygen-enriched shock as well. However, substantially more powerful broad magnetic field maxima are superimposed on these small-scale perturbations. The latter have obviously longer wavelengths, and their period appears to be about $150 \Omega^{-1}$, i.e. roughly $A/Z = 16$ times greater. This again reveals the resonant nature of the reformation process. The greater amplitudes of oxygen-generated fluctuations suggest that the weakly charged heavy ions start to govern the overall shock dynamics in case of their at least 50\% relative density (which means only about 6\% number abundance).

\begin{figure}[h]
\includegraphics[width=0.9\linewidth]{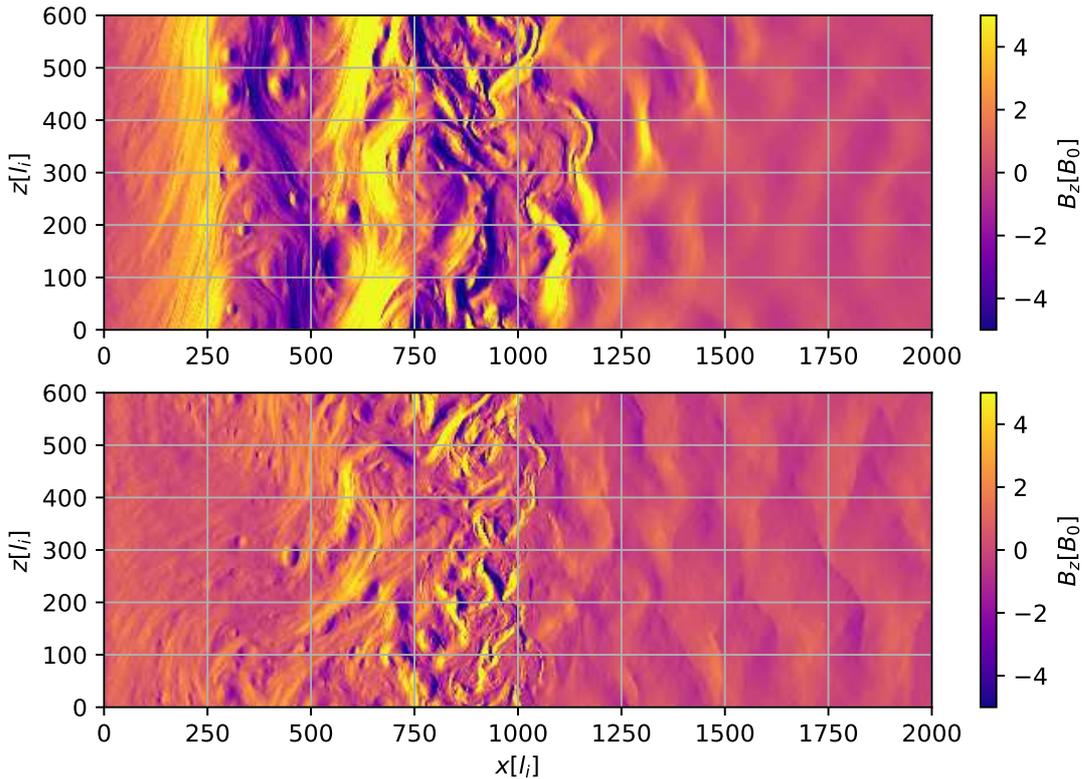}
\caption{The colormap of $B_z$ at $t=400\Omega^{-1}$  for the shock with 50\% O(+1) admixture (top panel) and pure hydrogen shock (bottom panel)}
\label{two_maps}
\end{figure}

Figure \ref{two_maps} illustrates the spatial structure of hydrogen and oxygen-dominated shocks. Again, the qualitative similarity of the upstream and downstream magnetic field structures can be seen together with the difference of spatial scales and amplitudes. The fourier analyses also showed powerful long-wave harmonics in the case of 50\% O(+1). 
It should be noted that the ratio of characteristic wavenumbers for the investigated shocks is less than oxygen $A/Z$. However,
the spatial scales of oxygen-dominated shock are likely to be affected by the limited sizes of simulation box, so  more resource-intensive setups are needed for the quantitative analyses. 

Simulations of shocks with $\theta = 0^o$ and $\theta  = 30^o$ showed that the reformation period tends to slightly increase with inclination angle. However, the difference is within temporal variations and cannot be considered significant. Meanwhile the qualitative picture of increasing characteristic scales for oxygen-enriched shocks remains the same for other inclinations.
It should be noted that Caprioli et al \cite{Caprioli2015} found smaller reformation period  (about $3 \Omega^{-1}$) for the shock with $M_a=20, \; \theta = 0^o$ and $\beta = 1$. The discrepancy is likely to originate from different upstream parameters (for example, the short-wavelength Bell instability plays the increasing role for higher Mach numbers). However, investigation of hydrogen shocks reformation for a wide parameter space is out of scope of this paper.

\section{Conclusions}
The supercritical collisionless shocks in cold magnetized plasma are found to experience the self-reformation process in both cases of the pure protons and the substantial O(+1) admixture. The temporal scales of the reformation process are proportional to the governing ions gyrofrequencies, thus demonstrating the resonant nature of the corresponding plasma instabilities. The oxygen-generated waves at shock appear to be more powerful than the hydrogen ones. Thus, even 
a 5-6\% (by number) admixture of weakly charged heavy ions takes over the control of the shock dynamics. 

Meanwhile velocity of the upstream and downstream scattering centers (i.e. wave fronts) does not depend on plasma composition and agrees well with analytical predictions. The upstream waves propagate against the incoming flow with the Alfv\'en velocity, while the downstream ones are at rest in the plasma frame.
The simulation results are applicable for the supernovae reverse shocks.

\section*{Acknowledgements}
Authors are grateful to Peter the Great St. Petersburg Polytechnic University for Tornado cluster access.
\newline
\bibliographystyle{iopart-num}
\bibliography{abs}{}

\end{document}